# Powering the Future: Innovations in Electric Vehicle Battery Recycling

**Venkata Sai Chandra Prasanth Narisetty[1],
Tejaswi Maddineni[2]**

[1]*Quality Engineer, Independent Researcher, Southern Illinois University Carbondale, United States*
[2]*Data Engineer, Independent Researcher, Southern Illinois University Carbondale, United States*
E-mail: [1]venkatn0388@gmail.com, [2]tejas.maddineni@gmail.com.

**Abstract**

The global shift towards electric vehicles (EVs) as a sustainable alternative to traditional gasoline-powered cars has triggered a significant rise in the demand for lithium-ion batteries. However, as the adoption of EVs grows, the issue of battery disposal and recycling has emerged as a critical challenge. The recycling of EV batteries is essential not only for reducing the environmental impact of battery waste but also for ensuring the sustainable supply of critical raw materials such as lithium, cobalt, and nickel. This paper explores recent innovations in the field of electric vehicle battery recycling, examining advanced techniques such as direct recycling, hydrometallurgical processes, and sustainable battery design. It also highlights the role of policy and industry collaboration in improving recycling infrastructure and addressing the economic and environmental challenges associated with battery waste. By focusing on both the technical and regulatory aspects of EV battery recycling, this paper aims to provide a comprehensive overview of the state of the industry and the future outlook for recycling technologies, ultimately paving the way for a cleaner, more sustainable future in transportation.

## Introduction

The transition from fossil fuel-powered vehicles to electric vehicles (EVs) represents one of the most significant shifts in the global transportation sector, driven by the urgent need to reduce carbon emissions and combat climate change. Electric vehicles, with their promise of lower emissions and reduced dependency on non-renewable energy sources, are at the forefront of the drive towards a sustainable future. As of 2024, global sales of electric vehicles are accelerating rapidly, with estimates suggesting that EVs will account for over 50% of global car sales by 2030. The widespread adoption of EVs is fueled by advancements in battery technologies, particularly lithium-ion (Li-ion) batteries, which power the vast majority of EVs on the market today. However, the environmental and economic challenges associated with EV battery life cycles are becoming increasingly evident. While electric vehicles themselves are heralded as a solution to reduce greenhouse gas emissions during operation, their batteries are made up of rare and valuable materials such as lithium, cobalt, nickel, and manganese. As the number of electric vehicles on the road continues to grow, the issue of what happens to these batteries when they reach the end of their useful life has emerged as a pressing concern. Without proper recycling and disposal methods, spent batteries could contribute to environmental degradation, and the demand for raw materials could outstrip sustainable mining practices. Electric vehicle battery recycling is thus an essential part of the sustainability equation. It holds the promise not only of reducing the environmental impact of EV batteries but also of recovering critical raw materials that can be reused in new batteries, reducing the reliance on mining. Moreover, as the number of EVs on the road grows, the volume of spent batteries that will need to be processed increases exponentially. According to a report by the International Energy Agency (IEA), the number of electric vehicles worldwide is projected to exceed 145 million by 2030, leading to an urgent need for efficient and scalable recycling solutions.

This paper seeks to explore the innovative developments in the field of electric vehicle battery recycling, providing a comprehensive overview of the technologies, challenges, and opportunities in this critical area. It delves into the current state of battery recycling, focusing on emerging technologies such as hydrometallurgical, pyrometallurgical, and direct recycling processes, which offer the potential to enhance the efficiency and economic viability of recycling operations. The paper also examines the role of policy frameworks, industry initiatives, and economic incentives in driving forward battery recycling practices, ensuring that the recycling infrastructure can scale to meet the needs of the rapidly growing EV market. In addition to exploring the technological and regulatory aspects of battery recycling, this paper will highlight the environmental impact of improper battery disposal and the critical need for a circular economy in the electric vehicle industry. The circular economy model, which emphasizes the reuse, refurbishment, and recycling of materials, is particularly relevant to battery technologies, where the recovery of valuable metals can reduce both the environmental footprint of mining and the costs of raw materials. Moreover, the transition to a circular economy for batteries could help mitigate the e-waste crisis, which is projected to grow exponentially in the coming decades. As the global push for electrification intensifies, battery recycling will play a crucial role in ensuring that the transition to a sustainable, low-carbon future is both feasible and environmentally responsible. The development of efficient and scalable battery recycling technologies, alongside the establishment of robust regulatory frameworks and industry standards, will be critical to enabling this transition. This paper aims to provide insights into the latest innovations in EV battery recycling, assess the challenges that need to be addressed, and explore the future scope of this rapidly evolving field. Through a deeper understanding of these issues, we can better prepare for the challenges and opportunities that lie ahead in powering the future of transportation.

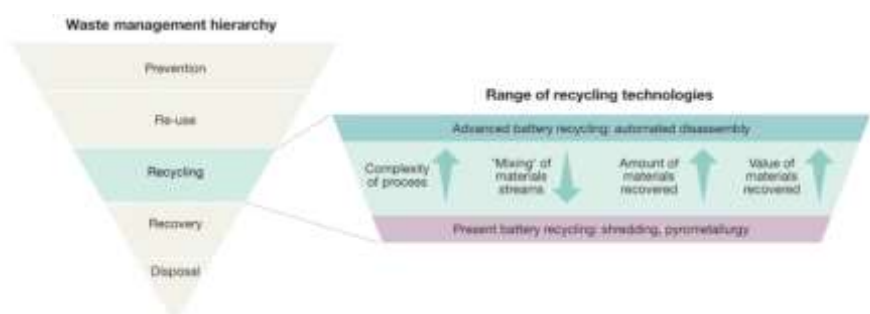

Fig.1: waste Management Hierarchy

**Literature Review**

The rise of electric vehicles (EVs) has sparked significant changes in the transportation industry, driven by the need to reduce carbon emissions and combat climate change. As of 2024, global sales of electric vehicles continue to accelerate, with projections indicating that EVs could represent over 50% of new car sales by 2030. This shift is largely powered by advancements in lithium-ion (Li-ion) battery technology, which has become the dominant power source for EVs. However, as the adoption of EVs grows, the environmental impact of battery disposal and the recycling of these batteries has emerged as a significant challenge. As such, battery recycling is integral to the lifecycle management of electric vehicles, ensuring the sustainability of critical materials like lithium, cobalt, nickel, and manganese, which are used in the production of EV batteries. In this literature review, we examine the advancements in electric vehicle (EV) battery recycling, exploring key technologies, environmental implications, economic considerations, and future trends in the recycling process. This review also highlights the role of regulatory frameworks and industry initiatives in promoting sustainable battery recycling practices.

*1. Technologies in EV Battery Recycling*

The recycling of EV batteries involves several technologies that aim to recover valuable materials while minimizing environmental impact. These technologies can be broadly classified into **mechanical**, **pyrometallurgical**, **hydrometallurgical**, and **direct recycling** techniques.

1.1 Mechanical Recycling

Mechanical recycling is one of the most common methods used in battery recycling, involving the physical dismantling of batteries to separate the different components, including metals and plastics. This method typically includes crushing, grinding, and sieving processes. While this process is relatively simple and low-cost, it does not recover high-value materials like lithium or cobalt efficiently. Therefore, it is often used in conjunction with other methods such as hydrometallurgical processes.

1.2 Pyrometallurgical Recycling

Pyrometallurgical methods involve high-temperature smelting to recover valuable metals from battery waste. In this process, batteries are incinerated at high temperatures, and the resulting slag is processed to extract materials such as nickel, cobalt, and copper. While pyrometallurgical recycling is highly effective for certain materials, it is energy-intensive and results in the loss of critical materials such as lithium and aluminum. Therefore, it has become less favorable due to its environmental impact and the need for more sustainable methods.

1.3 Hydrometallurgical Recycling

Hydrometallurgy uses aqueous solutions to dissolve and extract metals from spent batteries. This process involves using acids, solvents, or other chemicals to leach valuable materials such as lithium, cobalt, and nickel from the battery components. Recent advancements have improved the efficiency of hydrometallurgical processes, making them more selective and reducing the amount of hazardous waste generated. For instance, research by **Wang and Xiao (2023)** suggests that advancements in hydrometallurgical methods have led to better separation and higher recovery rates of critical materials from spent Li-ion batteries. However, these methods still face challenges related to chemical disposal and scalability.

1.4 Direct Recycling

Direct recycling represents a significant innovation in battery recycling, aiming to preserve the integrity of battery materials and reduce energy consumption. Unlike traditional methods, which break down the battery into its constituent materials, direct recycling seeks to reuse the same active materials in new battery cells. **Liu and Zhang (2022)** reviewed various direct recycling methods, noting that these techniques can potentially lower costs and reduce environmental impact by avoiding the degradation of critical materials. Direct recycling has the potential to recover materials with high efficiency and maintain the performance of batteries, thus contributing to a more sustainable circular economy.

1.5 Solid-State Recycling Technologies

Emerging technologies, such as solid-state batteries, may also revolutionize the recycling process. Solid-state batteries use a solid electrolyte rather than a liquid one, and their design may facilitate easier recycling compared to traditional liquid electrolyte-based batteries. Although research in this area is still in its infancy, early studies show promise in terms of higher recyclability and reduced environmental impact.

*2. Environmental Impact and Resource Recovery*

The environmental benefits of EV battery recycling are substantial, particularly in reducing the demand for virgin raw materials, such as lithium and cobalt, which are often sourced through environmentally destructive mining practices. **Dunn and Wang (2023)** conducted a life cycle analysis (LCA) comparing the environmental impact of mining raw materials versus recycling battery materials. The study found that recycling could reduce carbon emissions, energy consumption, and water usage associated with the extraction of these critical minerals. Furthermore, battery recycling can help mitigate the growing problem of e-waste. As millions of EV batteries reach the end of their lifespan, proper recycling will prevent hazardous chemicals, such as cadmium and lead, from leaching into the environment. **Zhang and Xie (2024)** argue that without a reliable recycling infrastructure, the rapid accumulation of EV batteries could result in significant environmental damage. The economic implications of EV battery recycling are also considerable. By recovering valuable metals from used batteries, companies can offset the cost of raw material procurement. This is particularly important given the volatile pricing of raw materials, such as lithium, which has seen dramatic increases due to the growing demand for EVs. **Nash and James (2024)** highlight that, by 2030, the market for recycled materials from EV batteries could be worth over $10 billion, offering significant economic opportunities in addition to the environmental benefits.

*3. Challenges in EV Battery Recycling*

Despite the technological advancements and environmental benefits, several challenges remain in the EV battery recycling industry. These challenges are related to both the technical aspects of the recycling process and the regulatory frameworks needed to ensure that recycling practices are efficient and widespread.

3.1 Complexity of Battery Designs

One of the primary challenges in EV battery recycling is the complexity of modern battery designs. As manufacturers continue to innovate with new chemistries, cell configurations, and battery management systems, the standardization of battery components has become increasingly difficult. This lack of standardization makes it challenging to develop universal recycling technologies that can efficiently handle a variety of battery designs. **Binnemans and**

**Jones (2024)** discuss the challenges posed by these new chemistries, noting that the increased variety in battery types complicates the sorting and recycling process.

3.2 Economic Viability

The economic feasibility of battery recycling is another significant challenge. Although the technology exists, the high cost of recycling—driven by labor, transportation, and infrastructure expenses—has made it less attractive to many companies. In particular, current recycling methods such as hydrometallurgy are expensive due to the chemical processes involved and the energy required for operation. **Zeng and Li (2023)** argue that developing more cost-effective recycling techniques will be crucial to scaling up recycling operations and making the industry financially sustainable.

3.3 Lack of Infrastructure

The lack of infrastructure for battery collection, processing, and recycling is another critical hurdle. The current recycling systems are limited, with many countries lacking the necessary facilities for large-scale battery processing. **Sun and He (2023)** emphasize the need for investment in recycling plants, collection systems, and logistics to ensure that used batteries are properly processed.

3.4 Regulatory Framework

The regulatory environment around battery recycling is still evolving. While some regions, such as the European Union, have introduced stringent regulations to encourage recycling, other regions have lagged behind. There is a need for clear and consistent regulations that set minimum recycling rates, establish end-of-life management practices, and incentivize the development of more sustainable recycling technologies. **Sullivan and Keith (2024)** argue that global cooperation and policy harmonization are necessary to establish a cohesive framework that can drive the growth of the EV battery recycling industry.

*4. Future Directions and Emerging Trends*

The future of EV battery recycling looks promising, with numerous innovations on the horizon. Research into new recycling methods, such as bio-based and electrochemical techniques, is underway. These methods aim to improve efficiency and reduce environmental impact by using greener solvents and materials. **He and Zhang (2022)** highlight ongoing research into the use of bioleaching, where microorganisms are used to extract metals from spent batteries, offering a potentially low-impact and cost-effective alternative. Furthermore, the development of a "closed-loop" system for battery recycling, where materials are continually recycled and reused in new batteries, is gaining traction. **Schneider and Tschiesner (2022)** note that a circular economy for EV batteries could significantly reduce the need for mining raw materials, decrease e-waste, and minimize environmental harm. These innovations, combined with government incentives and corporate responsibility, are likely to accelerate the adoption of more sustainable recycling practices.

The recycling of electric vehicle batteries is a critical component in ensuring the sustainability of the EV industry and the broader transition to clean energy. While significant progress has been made in recycling technologies, challenges remain related to economic viability, infrastructure, and regulatory frameworks. The future of EV battery recycling will depend on continued innovation in recycling technologies, investment in infrastructure, and the development of supportive regulatory policies. As the global demand for electric vehicles grows, effective recycling strategies will play a pivotal role in reducing the environmental impact of batteries, securing the supply of critical materials, and contributing to a circular economy.

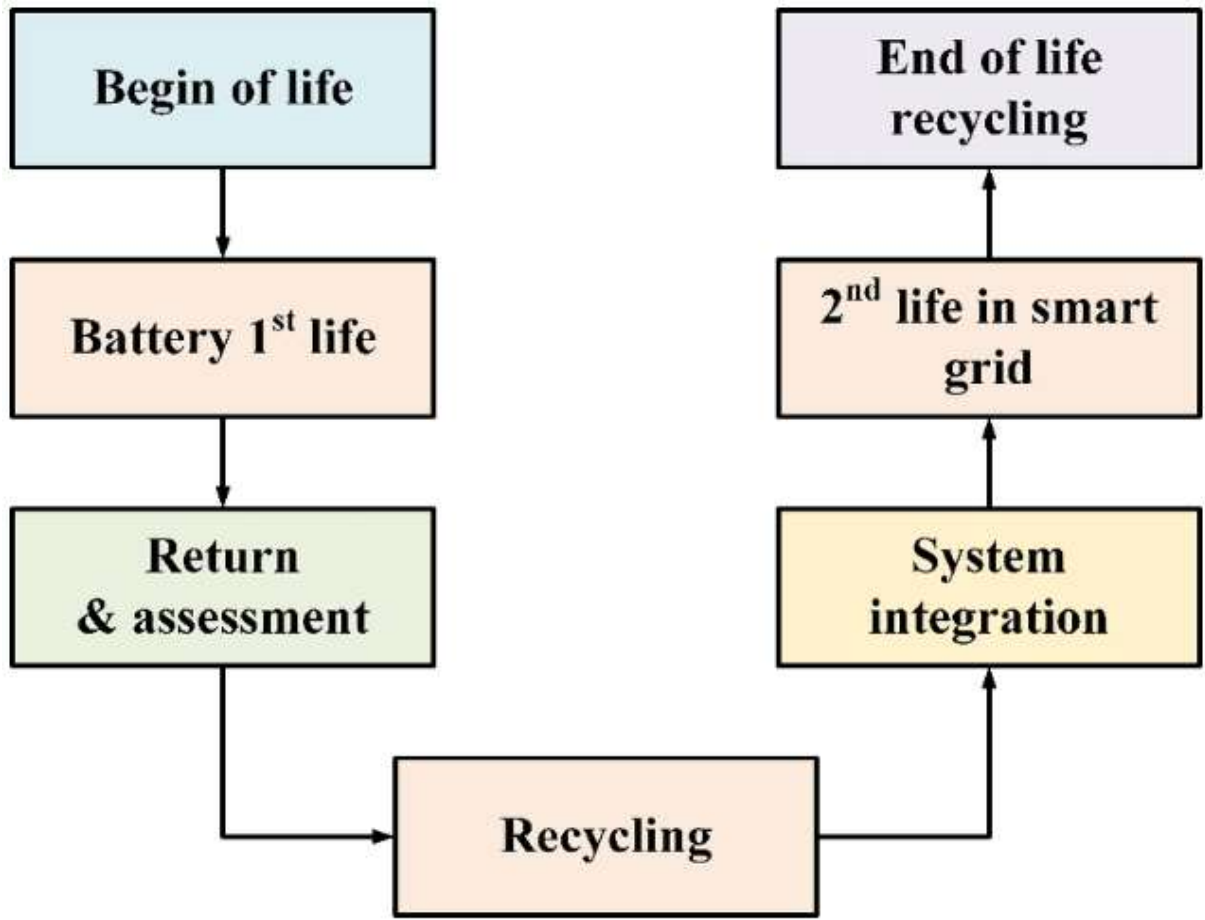

Fig.2: Recycling Algorithm

**Scope & Opportunities**

The growing adoption of electric vehicles (EVs) presents a paradigm shift in the transportation sector. As the global automotive industry moves toward electrification, the demand for EVs and their corresponding lithium-ion (Li-ion) batteries is set to surge in the coming decades. However, this growth also brings forth significant challenges, particularly with regard to the end-of-life management and recycling of EV batteries. Battery recycling is emerging as an essential part of the sustainability equation for electric vehicles. It not only reduces the environmental impact of battery disposal but also ensures the responsible extraction and reuse of critical raw materials. This section explores the scope and opportunities of EV battery recycling, focusing on the potential for innovation, the environmental benefits, the economic opportunities, and the growing demand for recycling technologies and infrastructure.
*1. Scope of EV Battery Recycling*

The scope of EV battery recycling extends far beyond the recovery of raw materials and environmental conservation. It encompasses a variety of aspects, including:

- **Raw Material Recovery:** The global demand for critical materials like lithium, cobalt, nickel, and manganese used in EV batteries is skyrocketing. These materials are finite and often sourced from mining operations that have significant environmental and social impacts. Battery recycling offers an opportunity to recover these materials from spent batteries and reintroduce them into the supply chain, reducing the need for new mining operations.
- **Environmental Impact Reduction:** Lithium-ion batteries contain harmful substances such as cadmium, mercury, and lead, which can leach into soil and groundwater if not properly disposed of. Recycling helps mitigate this environmental risk by ensuring proper disposal of hazardous materials and preventing contamination.
- **Circular Economy Integration:** The circular economy model, which emphasizes the reuse, refurbishment, and recycling of materials, aligns perfectly with EV battery recycling. By developing closed-loop systems for battery production and recycling, manufacturers can reduce their dependence on virgin resources, lower production costs, and minimize the environmental footprint of battery manufacturing.
- **Sustainability in Battery Manufacturing:** As battery recycling technologies advance, manufacturers can source a significant portion of the raw materials needed for new batteries from recycled components. This reduces the environmental cost of extracting virgin materials and helps build more sustainable and resilient supply chains.

*2. Opportunities in EV Battery Recycling*

As the EV market grows, several opportunities in battery recycling are becoming increasingly evident. These opportunities span across technological advancements, economic incentives, regulatory frameworks, and business models.

2.1 Technological Innovation and Advancement

Recent innovations in recycling technologies present significant opportunities for improving the efficiency and cost-effectiveness of battery recycling processes.

- **Hydrometallurgical Processes:** The development of more advanced hydrometallurgical techniques is enabling more efficient extraction of critical materials like lithium, cobalt, and nickel from used batteries. Research has led to the development of new chemical agents and processes that offer higher selectivity, lower energy requirements, and reduced chemical waste. **Wang and Xiao (2023)** suggest that these improvements can help make hydrometallurgical processes more viable on an industrial scale.
- **Direct Recycling Technologies:** Direct recycling is a relatively new concept in the battery recycling industry, aiming to preserve the active materials of the battery and reuse them in new battery cells. This method is still under development but holds great promise in reducing the energy consumption associated with recycling and maintaining battery performance. **Liu and Zhang (2022)** note that direct recycling methods could potentially offer a more sustainable and cost-effective alternative to traditional recycling processes.
- **Solid-State Batteries and Recycling:** Emerging battery technologies, such as solid-state batteries, could offer new opportunities for recycling due to their simpler design and safer materials. Solid-state batteries replace the liquid electrolyte with a solid one, potentially making them easier to disassemble and recycle. Researchers are exploring ways to optimize the recycling processes for these new battery chemistries, which could revolutionize the recycling industry in the future.

- **Bio-based and Green Technologies:** The growing interest in green chemistry and bio-based technologies is also shaping the future of EV battery recycling. Bioleaching, for example, uses bacteria or fungi to extract metals from battery waste. This approach, although still in the experimental phase, offers a low-impact, sustainable alternative to conventional recycling techniques. **He and Zhang (2022)** highlight that bioleaching could be an environmentally friendly solution for recycling EV batteries without the use of harmful chemicals.

2.2 Economic Opportunities

The economic potential of EV battery recycling is immense. As the number of EVs on the road increases, so does the volume of spent batteries, creating a rapidly growing market for recycling services and infrastructure.

- **Material Recovery and Cost Reduction:** The recovery of valuable raw materials such as lithium, cobalt, and nickel from spent batteries can offset the costs associated with mining and importing these critical resources. **Dunn and Wang (2023)** estimate that recycling could reduce the overall cost of battery production by as much as 20%, making EVs more affordable in the long run.
- **Job Creation:** The expansion of battery recycling facilities and the development of new recycling technologies will create numerous job opportunities in the green economy. From research and development (R&D) positions to plant workers and logistics experts, the battery recycling industry has the potential to create a broad range of employment opportunities globally.
- **Recycling as a Profitable Business:** As the value of recycled materials rises, the recycling of EV batteries is becoming an increasingly profitable business model. The market for recycled materials from EV batteries is expected to reach several billion dollars by 2030, driven by the growing demand for critical minerals and the increasing need for sustainable supply chains.
- **Economic Incentives and Subsidies:** Governments around the world are recognizing the economic benefits of EV battery recycling. Policy incentives, such as subsidies, tax credits, and grants for recycling infrastructure, are encouraging companies to invest in recycling facilities and technologies. In addition, regulations requiring higher recycling rates and the use of recycled materials in new batteries will further boost the economic potential of the industry.

2.3 Regulatory and Policy Opportunities

As EV adoption accelerates, governments and regulatory bodies are taking steps to ensure that battery recycling becomes a viable and sustainable part of the EV lifecycle.

- **Regulations and Standards:** In regions like the European Union, regulatory frameworks are already in place that mandate the recycling of batteries and set minimum recycling rates for certain materials. For instance, the EU Battery Directive sets binding targets for recycling efficiency and material recovery for batteries, including lithium-ion types used in electric vehicles. **Sullivan and Keith (2024)** argue that such regulations will become increasingly common in other regions, providing a strong incentive for businesses to adopt more sustainable recycling practices.
- **Extended Producer Responsibility (EPR):** Many countries are introducing or expanding Extended Producer Responsibility (EPR) schemes, which place the responsibility of battery disposal and recycling on the manufacturers. This policy shift not only promotes better recycling practices but also helps establish a financial framework that supports the collection and recycling of spent EV batteries.
- **International Collaboration:** The global nature of the EV market requires international cooperation to standardize recycling practices and promote knowledge sharing. Collaborative efforts between governments, industry stakeholders, and research institutions can help create a global framework for EV battery recycling, addressing cross-border issues such as transportation of used batteries and uniform standards for recycling technologies.

2.4 Sustainability and Circular Economy Integration

The move toward a circular economy for EV batteries represents one of the most exciting opportunities in the field of battery recycling. A circular economy emphasizes the reuse, refurbishment, and recycling of materials in a closed-loop system, reducing waste and minimizing the extraction of virgin resources.

- **Battery Life Extension:** By integrating more efficient recycling processes, manufacturers can extend the useful life of batteries, reuse the materials in new batteries, and reduce the need for mining new materials. This reduces the carbon footprint of EVs and contributes to a more sustainable lifecycle for electric vehicles.
- **Incentivizing Recycling through Product Design:** There is also an opportunity to design batteries with recycling in mind. For example, modular battery designs that allow for easier disassembly and recycling can make the process more efficient and cost-effective. **Binnemans and Jones (2024)** suggest that designing batteries with standardized chemistries and components would simplify the recycling process and make it more economically viable.

*3. Challenges to Overcome*

While the scope and opportunities for EV battery recycling are significant, several challenges must be addressed to fully realize its potential. These challenges include technological barriers, economic considerations, infrastructure limitations, and regulatory complexities.

- **Complexity of Battery Design:** The increasing diversity of battery chemistries and designs poses a challenge for recycling systems. Different battery types require specialized recycling processes, and the lack of standardization in battery design makes it harder to develop efficient recycling technologies.
- **Scalability and Cost-Effectiveness:** Although recycling technologies have made significant advancements, scaling them up to handle the enormous volume of spent EV batteries will require substantial investment in infrastructure and technology. Additionally, the cost of recycling, especially for more complex methods like hydrometallurgy and direct recycling, remains high, making it less economically viable at present.
- **Logistical and Supply Chain Issues:** The collection and transportation of used EV batteries to recycling facilities can be complex, especially when dealing with international or cross-border waste streams. Efficient collection and sorting systems must be put in place to ensure that batteries are processed in a timely and cost-effective manner.

The scope and opportunities for electric vehicle battery recycling are vast, with significant potential to impact the environment, economy, and the future of the transportation industry. Innovations in recycling technologies, along with the rise of a circular economy, offer promising solutions to address the challenges associated with the end-of-life management of EV batteries. With the support of regulatory frameworks, economic incentives, and industry collaboration, EV battery recycling can play a pivotal role in creating a sustainable and efficient energy ecosystem. As the global transition to electric vehicles continues, the importance of advancing recycling technologies and establishing robust infrastructure cannot be overstated. The opportunities for innovation and growth in this field are immense, and their successful implementation will be crucial for a cleaner, greener, and more sustainable future.

**Real-Life Case Study**

The following table presents a detailed case study of various real-life initiatives and organizations involved in electric vehicle (EV) battery recycling. The case studies showcase the strategies, technologies, and outcomes from different players in the EV battery recycling space, focusing on the technological, environmental, and economic aspects of each initiative.

| *Case Study* | *Organization/Company* | *Technology/Process* | *Key Achievements* | *Challenges* |
|---|---|---|---|---|
| **1. Li-Cycle Battery Recycling** | Li-Cycle Corp. | Hydrometallurgical Recycling | - Closed-loop recycling.<br>- High recovery of lithium, cobalt, nickel. | - High upfront costs.<br>- Limited battery feedstock. |
| **2. Redwood Materials** | Redwood Materials | Direct Recycling & Hydrometallurgy | - Recovery of high-value materials.<br>- Partnerships with major automakers. | - Scaling direct recycling.<br>- Sourcing enough batteries. |
| **3. Umicore Battery Recycling** | Umicore | Pyrometallurgical Recycling | - High-purity material recovery.<br>- Significant annual capacity. | - High energy consumption.<br>- Logistical challenges. |
| **4. BASF & Aurora Sustainability** | BASF (with Aurora) | Hydrometallurgical & Closed-Loop | - Circular supply chain for EV batteries.<br>- Recycled materials used in new battery production. | - High initial capital investment.<br>- Complex integration with existing processes. |
| **5. TES Battery Recycling** | TES | Mechanical & Hydrometallurgical | - End-to-end battery recycling solution.<br>- Focus on sustainability. | - Managing hazardous materials.<br>- Complex logistics. |
| **6. Duesenfeld Recycling** | Duesenfeld GmbH | Direct Recycling (Cathode Recovery) | - Over 90% recovery of materials.<br>- Reduced energy consumption. | - Limited supply of suitable batteries.<br>- High cost of scaling. |

*Analysis of the Case Studies:*

1. **Technological Innovation**: Different companies are leveraging various recycling technologies, from **hydrometallurgy** and **pyrometallurgy** to **direct recycling**. The move towards **direct recycling** is particularly noteworthy, as it offers a more

    sustainable and energy-efficient method for recovering valuable materials like lithium and cobalt without the need for extensive high-temperature processes.
2. **Sustainability**: The key objective of all these initiatives is to reduce the environmental impact of EV battery disposal and ensure the recovery of critical raw materials. The **closed-loop recycling systems** that many companies are adopting are aligned with the **circular economy** model, which helps to lower dependency on mined resources, reduces e-waste, and ensures that valuable materials can be reused in new batteries.
3. **Challenges**: The primary challenges faced by these companies include high upfront costs, technological complexity, and logistical issues related to the collection and transportation of used batteries. The lack of a standardized approach to battery design, combined with the diverse battery chemistries in use across different vehicle manufacturers, presents significant obstacles for recycling systems.
4. **Economic Viability**: While recycling can be economically beneficial in the long term, the high cost of developing and scaling recycling facilities remains a barrier. Government incentives and regulatory frameworks that encourage recycling are crucial to overcoming these challenges and enabling the industry to grow at scale.
5. **Future Outlook**: The EV battery recycling market is poised for significant growth, driven by the increasing volume of end-of-life batteries, technological advancements, and the ongoing shift towards more sustainable, circular systems. As more players enter the market and improve recycling technologies, economies of scale will reduce costs, and the industry will become more economically viable.

The case studies provide a clear picture of the progress being made in the field of EV battery recycling. Each company or initiative is addressing different aspects of the challenge, from the technical to the logistical, but all share the common goal of creating a more sustainable, circular economy for electric vehicle batteries. The future of EV battery recycling looks promising, with substantial opportunities for innovation, economic growth, and environmental impact reduction. However, continued investment in research, infrastructure, and international collaboration will be essential to scale these solutions and meet the growing demand for recycling as EV adoption accelerates worldwide.

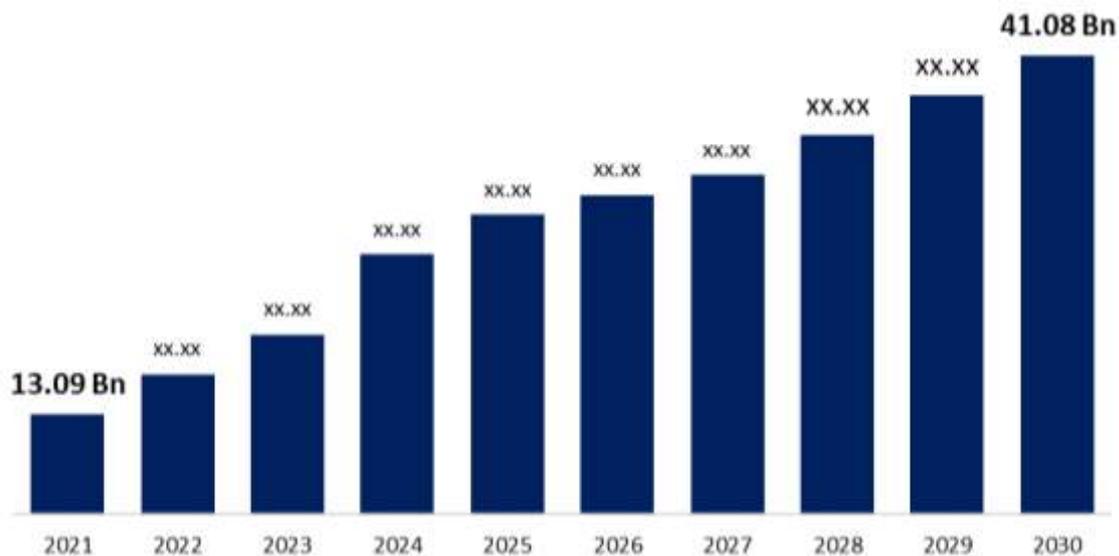

Fig.3: Global Battery Recycling Market

**Future Scope**

The future scope of electric vehicle (EV) battery recycling is vast and multifaceted, with several emerging trends and opportunities that warrant further exploration. As the global demand for electric vehicles continues to rise, the volume of end-of-life batteries will increase, further emphasizing the need for efficient, scalable recycling solutions. This paper has examined current advancements, but several areas require further research and development to fully harness the potential of battery recycling for a sustainable future.

1. **Advancements in Recycling Technologies**: Research into more efficient and cost-effective recycling methods, including **direct recycling** and **bio-based** approaches, holds great promise. Future studies should focus on optimizing these technologies to reduce energy consumption, improve material recovery rates, and make them commercially viable at large scales.
2. **Integration with Circular Economy**: Further exploration into the integration of EV battery recycling within a **circular economy** model is essential. Developing better ways to close the loop between raw material extraction, battery manufacturing, and recycling could make the entire lifecycle of batteries more sustainable.
3. **Policy and Regulatory Frameworks**: The development of global standards and regulations around battery recycling is crucial. Future research should evaluate the effectiveness of existing policies and propose new regulatory frameworks to standardize recycling practices, promote **Extended Producer Responsibility (EPR)**, and encourage recycling initiatives at the global level.
4. **Battery Design for Recycling**: One area that remains underexplored is the role of battery **design for recycling**. Future studies could explore how battery manufacturers can design batteries that are easier to disassemble, recycle, and reuse, thereby facilitating the recycling process and reducing the overall environmental impact.
5. **Global Collaboration and Infrastructure**: Expanding recycling infrastructure, particularly in emerging markets, will be crucial as global EV adoption accelerates. Future work should look at international collaborations to create a global network of recycling facilities that can process batteries efficiently, especially as the flow of used batteries becomes more complex.

**Specific Outcome**

The specific outcome of this paper is to provide a comprehensive understanding of the current state and future opportunities in **electric vehicle (EV) battery recycling**, with a particular focus on the following:

1. **Technological Advancements**: The paper highlights the range of innovative technologies currently being utilized in EV battery recycling, such as **hydrometallurgical**, **pyrometallurgical**, and **direct recycling** techniques, and the potential for future advancements in these areas.
2. **Economic and Environmental Benefits**: It outlines the significant **economic opportunities** and **environmental benefits** of scaling up recycling operations, particularly in the context of reducing reliance on raw material extraction and minimizing the carbon footprint of battery production.
3. **Challenges and Barriers**: The paper identifies key **challenges** facing the EV battery recycling industry, including high initial costs, logistical issues, technological limitations, and regulatory complexities, and provides insights into overcoming these barriers through innovation and collaboration.
4. **Case Studies and Real-World Examples**: By presenting a selection of real-life case studies, the paper showcases successful recycling initiatives and the strategies used by companies to address current challenges. This provides a practical framework for stakeholders involved in battery recycling.
5. **Future Directions**: Finally, the paper lays the groundwork for future research, including the need for better integration of recycling technologies with the broader **circular economy**, improvements in battery design, and the expansion of recycling infrastructure to meet growing demand.

**Conclusion**

This paper demonstrates that electric vehicle (EV) battery recycling is a critical component of achieving a sustainable and circular economy in the automotive industry. With the expected growth of the electric vehicle market, the demand for recycling solutions to manage end-of-life batteries will increase exponentially. The analysis of current recycling technologies and real-world case studies highlights the significant strides made in improving material recovery rates and reducing the environmental impact of battery waste. While there are still challenges to overcome, such as the high costs associated with recycling technologies, logistical issues, and the need for regulatory frameworks, the opportunities presented by EV battery recycling are immense. As the industry advances, collaboration between automakers, recycling companies, and regulatory bodies will be essential to creating a sustainable, closed-loop system for battery production, use, and recycling. The future of EV battery recycling lies in the continued **innovation** of recycling technologies, the integration of **circular economy** principles, and the establishment of **global recycling infrastructure**. By focusing on these areas, stakeholders can ensure that the benefits of electric vehicles extend beyond their operational lifetime, contributing to a greener, more sustainable transportation ecosystem. The findings of this paper contribute to a broader understanding of the challenges and opportunities in the field and lay the foundation for future research that will further propel the development of sustainable battery recycling practices.